\documentclass[journal,twoside,web]{ieeecolor}
\usepackage{tuffc}
\usepackage{biblatex}
\usepackage{amsmath,amssymb,amsfonts}
\usepackage{algorithmic}
\usepackage{graphicx}
\usepackage{textcomp}
\usepackage{wrapfig,colortbl}
\definecolor{abstractbg}{rgb}{1,0.969,0.914}
\def\BibTeX{{\rm B\kern-.05em{\sc i\kern-.025em b}\kern-.08em
    T\kern-.1667em\lower.7ex\hbox{E}\kern-.125emX}}

\newcommand{\bi}[1]{\boldsymbol{#1}}

\begin{document}

\title{AcousTools: A `Full-Stack', Python-Based, Acoustic Holography Library}

\author{Joshua~Mukherjee,
        Giorgos~Christopoulos,
        Zhouyang~Shen,
        Sriram~Subramanian,
        and Ryuji~Hirayama
\thanks{This work was supported by the UK Research and Innovation through their Frontier Research Guarantee Grant (No. EP/X019519/1), the Royal Academy of Engineering through their Chairs in Emerging Technology Program (No. CIET 17/18) and through the EPSRC prosperity partnership (No. EP/V037846/1). For the purpose of open access, the authors has applied a Creative Commons Attribution (CC BY) licence to any Author Accepted Manuscript version arising from this submission}
\thanks{J. Mukherjee, G. Christopoulos, Z. Shen, S. Subramanian and R. Hirayama are with University College London,  Gower St, London WC1E 6BT. e-mail: joshua.mukherjee.19@ucl.ac.uk}}%

\IEEEtitleabstractindextext{%
\fcolorbox{abstractbg}{abstractbg}{%
\begin{minipage}{\textwidth}\rightskip2em\leftskip\rightskip\bigskip
\begin{wrapfigure}[30]{r}{3in}%
\hspace{-3pc}\includegraphics[width=2.9in]{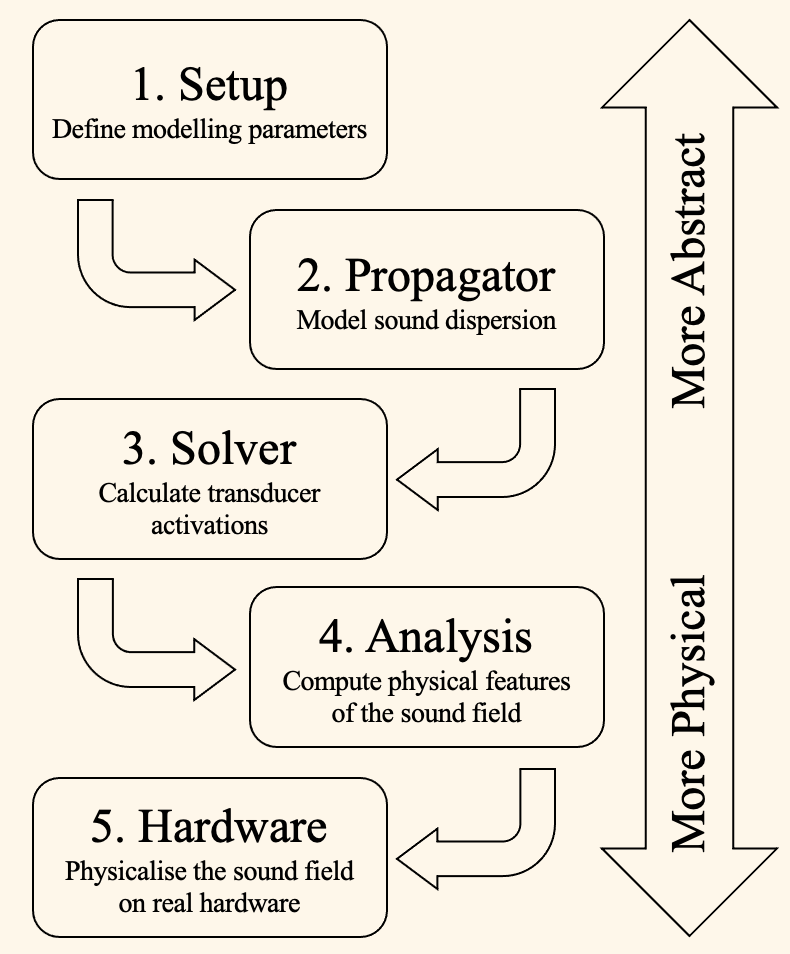}
\end{wrapfigure}%

\begin{abstract}
    Acoustic Holography is an emerging field where mid-air ultrasound is controlled and manipulated for novel and exciting applications. These range from mid-air haptics, volumetric displays, contactless fabrication, and even chemical and biomedical applications such as drug delivery. To develop these applications, a software framework to predict acoustic behaviour and simulating resulting effects, such as applied forces or scattering patterns is desirable. There have been various software libraries and platforms that attempt to fill this role, but there is yet to be a single piece of software that acts as a `full-stack' solution. We define this full-stack as the process from abstraction to physicalisation starting with setup, modelling acoustic propagation, transducer phase retrieval, sound field analysis, and control of the acoustic holographic hardware itself. Existing methods fail to fulfil one or more of these categories. To address this, we present AcousTools, a Python-based acoustic holography library, designed to support the full suite of acoustic holographic applications and we show AcousTools's ability to meet each step of the full-stack's requirements. AcousTools has the potential to become the standard code library for acoustic holography, with the uniquely complete suite of features wrapped in a language that is known to be easy to use, AcousTools will increase the ability for researchers to develop novel applications as well as accurately review other's work. The full-stack, aside from software, will also be useful for researchers -- providing a way to view and compare methodologies by understanding where they fit into the stack. 
\end{abstract}

\bigskip
\end{minipage}}}

\maketitle

\begin{figure*}
    \centering
    \includegraphics[width=\linewidth]{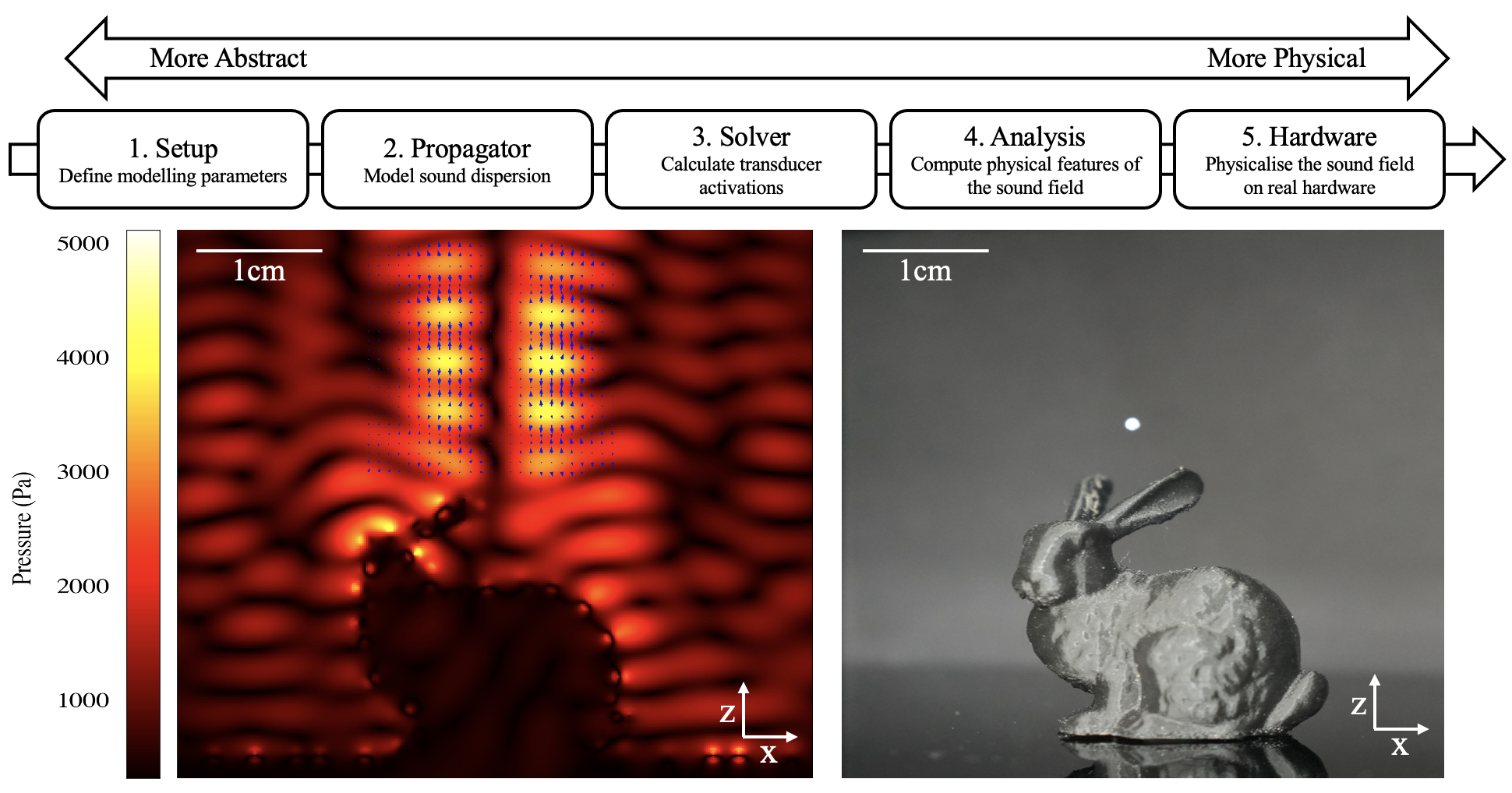}
    \caption{The acoustic holography full-stack (top) describes the workflow for acoustic holography. A developer begins on the left and progresses to the right, moving from the most abstract stages (defining modelling parameters and modelling propagation) to the most physical stages (examining physical metrics of a sound field and producing that sound field in reality). This framework allows the analysis of different acoustic holography software based on the stages they support -- the ideal software would allow a developer to complete every stage using a single software package. AcousTools uniquely addresses all stages in the full stack as shown in this example (bottom), from (1) defining the board and scatterer for levitation setup. Here a flat array above the working volume facing down towards a flat surface made of acrylic with an acrylic rabbit on top, together acting as a reflector. The array and reflector are separated by 12cm of air, (2) computing how the sound propagates with the scattering profile by using the boundary element method, (3) solving for the transducers' activation to focus sound waves above the rabbit's head while accounting for scattering, and then (4) analysing the acoustic force on a small particle (shown as blue arrows) before (5) finally rendering this in the real-world hardware. }
    \label{fig:headerFig}
    \label{fig:bunny_force}
    \label{fig:FullStack}
\end{figure*}

\section{Introduction}   

\begin{table*}[!t]
\arrayrulecolor{subsectioncolor}
\setlength{\arrayrulewidth}{1pt}
{\sffamily\bfseries\begin{tabular}{lp{6.75in}}\hline
\rowcolor{abstractbg}\multicolumn{2}{l}{\color{subsectioncolor}{\itshape
Highlights}{\Huge\strut}}\\
\rowcolor{abstractbg}$\bullet$ & Acoustic holography has neither a unified framework to think about acoustic applications nor a single software library to implement them --  reducing interoperability and flexibility.\\
\rowcolor{abstractbg}$\bullet${\large\strut} & We present the Acoustic Full-Stack, which describes the process a developer goes through to implement an acoustic application and then present AcousTools; a python-based library to implement each of these stages.\\
\rowcolor{abstractbg}$\bullet${\large\strut} & With a unified software library, the interpretability between researchers will increase and the full-suite of options means developers can make fully informed choices about which methods and algorithms to use.\\[2em]\hline
\end{tabular}}
\setlength{\arrayrulewidth}{0.4pt}
\arrayrulecolor{black}
\end{table*}

    \IEEEPARstart{N}{on-contact} manipulation and stimulus opens the door for numerous applications in a user-friendly and ergonomic way. Objects can be fabricated without ever touching the materials used, either by assembling smaller building blocks \cite{LeviPrint} or by shaping raw materials \cite{AcoustoFab}. Haptic stimulus can give greater immersion in virtual reality \cite{HapticVR}, leveraging human tactile mechanisms to produce higher quality stimulus \cite{HapticsMPSTM, HapticIllusions} and render complex stimulus such as softness \cite{HapticSoftness}. Levitation of small particles, in addition to fabrication, allows the creation of computer displays that are truly 3D \cite{VolDisplaySingleParticle}, by using props to represent objects \cite{Articulev} or by exploiting persistence of vision (PoV) to render shapes in 3D \cite{GSPAT, VolDisplaySingleParticle}. All these applications and more are unlocked by utilising focused mid-air ultrasound.
    \par
    To create these applications, a software package or library that can model the behaviour of acoustic waves is advantageous \cite{OpenMPD, Ultraino}. Common tasks include computing phases for transducer arrays in complex scenarios \cite{LeviPrint}, predicting how sound scatters from irregular objects \cite{HighSpeedBEM}, and analysing the force exerted on an object in a sound field \cite{AcousticElements}. Pre-built software allows developers to focus on higher-level goals without needing to implement these complex processes from scratch. 
    \par
    To compare software implementations for these tasks, we propose the acoustic `full-stack', a conceptual framework for developing acoustic applications that outlines the process a developer follows from concept to physical deployment. This framework has five stages as shown in Fig.\ref{fig:FullStack}. It begins with \textbf{Setup}, where the developer defines parameters for their application, such as the transducer arrays to be used and the locations that sound should be focused at. Next, \textbf{Propagators} is where the developer selects a method for modelling the acoustic waves, largely choosing between free-field assumptions or models that consider scattering effects. Then, \textbf{Solvers} is where the developer selects algorithms to compute the signals to be sent to the transducer arrays, with different algorithms having different advantages and trade-offs. In \textbf{Analysis}, the developer uses different metrics to compute the quality of the resulting sound field, either for quantification or to inform refinement of earlier stages. Finally, \textbf{Hardware} is where the computed signals are sent to physical devices, allowing the applications to be rendered in the real world.
    \par
    There are a variety of software packages that attempt to meet the stages described in the full-stack \cite{OpenMPD, Ultraino, Levitate, Kwave, FOCUS}. However, all existing solutions fail to meet one or many of these categories as can be seen in Tab.\ref{tab:software_summary}. As a result, developers often need to switch between different packages or are unable to use certain methodologies without implementing them independently. 
    \par
    As a unified solution for developing applications based on acoustic holography, we present AcousTools, a python-based library that can support every stage of the acoustic `full-stack'. Elements of this `full-stack' are shown in Fig.\ref{fig:bunny_force}. In the Setup stage, an array above a rabbit reflector is defined which means that for the Propagators phase the chosen propagator considers scattered contributions from the rabbit. Using this scattering, the waves can be focused at a point above the rabbit's head in the Solvers step and the resulting force is computed in the Analysis phase. Finally, in the Hardware step the field is rendered in reality above a 3D printed rabbit and a particle is levitated at the point that is predicted by the converging forces.
    \par
    In this work, we step through each stage of the full-stack, showing AcousTools' efficacy in that area and highlighting where other software falls short. Its implementation using PyTorch in Python leverages the language's ease of access especially for beginners \cite{pythonEase}, as well as the parallelised and comprehensive capabilities of the PyTorch environment \cite{PyTorch}. AcousTools is therefore the general purpose software solution that the acoustic holography community currently lacks. We anticipate that the community will adopt it as a standard code library, leading to a greater degree of interoperability and reproducibility among researchers.
    \par
    Additionally, we believe that the full-stack itself will give researchers a reliable mental model when considering acoustic propagators, algorithms, and applications by clarifying where their work sits in the wider ecosystem; supporting more informed comparisons with other methodologies. By offering an overall structure to the field of acoustic holography, the framework encourages more systematic thinking about research contributions.

\section{Related Work}

\subsection{Acoustic Holography}
    Acoustic holography, used to produce 3D sound fields \cite{HoloForAcoustics, AcousticElements, GSPAT, HighSpeedBEM}, can be used to focus high frequency sound waves. This allows for non-contactless manipulation \cite{VolDisplaySingleParticle, LeviPrint} and stimulation \cite{eigensolver, HapticsMPSTM} of materials at a distance. Unlike magnetic levitation, it is not limited to ferrous materials, and unlike optical holography using lasers, there is no dangerous heating of the samples \cite{2DDroplets}. Thus, acoustic holography is an exciting and interactive method. Acoustic levitation is even used in the field of nanoparticle mid-air chemistry \cite{AcousticChemistry}, but only the simplest acoustic setups \cite{TinyLev} are used in these cases due to the complexity of the control software commonly used in holographic work. Modern open-source hardware \cite{Ultraino, OpenMPD} combined with reliable and flexible software has a potential to expand the domains where acoustic holography can be used.

    \par
    
    \textbf{Propagation Methods: }Acoustic holography computes a set of transducer activations used to produce a desired 3D sound field \cite{AcousticElements}. It predicts how these waves interfere and build up \cite{VolDisplaySingleParticle, GSPAT, Diff-PAT} by modelling sound propagation. Sound waves are emitted from a source, often an array of transducers, and typically modelled using the Piston Model \cite{GSPAT, LeviPrint, HapticsMPSTM, VolDisplaySingleParticle, AcoustoFab}. In this model, the propagation from a single transducer $t \in \bi \tau$, for a set of transducers $\bi \tau$, located at $\bi z_t$ to a point in space located at $\bi z_n$ can be modelled as \cite{GSPAT}
        \begin{equation}
            F_{n, t} = \frac{2p_{ref}J_1(kr\sin(\theta))}{kr\sin(\theta)  d(\bi z_n, \bi z_t)}e^{ik  d(\bi z_n, \bi z_t)}. 
            \label{eq:Ftz}
        \end{equation}
        Here, $p_{ref}$ is the reference pressure at 1m for the transducer, $\theta$ is the angle between the transducer normal and the point, $r$ is the transducer radius, $k$ is the wavenumber, $J_1$ is the Bessel function of the first kind, and $d$ is the distance from the transducer and the point. The actual complex pressure can be then found by multiplying $F_{n,t}$ by the activation of that transducer, $x_t = A_t e^{i \varphi_t}$ where $A_t$ and $\varphi_t$ are the amplitude and phase, respectively. The values of $F_{n,t}$ and $x_t$ for each transducer and point can be combined into matrix forms, and the overall acoustic pressure field at a set of $N$ points, $\bi z_n \in \bi \nu, |\bi \nu| = N$, is given by 
        \begin{equation}
            \bi p_{\bi \nu} = \bi F_{\bi \nu \bi \tau}\ \bi x_{\bi \tau}.
            \label{eq:pressure}
        \end{equation}
    In a free-field scenario without scattering or occlusion, this model is sufficient. However, considering scattering would provide benefits to many applications. When there are objects in a levitation domain, for example when an object is fabricated during levitation, a hand is present when rendering haptics or a barrier blocks a levitation path, modelling sound scattering can increase manipulation precision and reduce potential levitation failure \cite{AcoustoFab}. This scattering media can be any sound-hard material, such as plastics or the surface of water and as many common materials have an acoustic impedance much higher than that of air, most materials act as scatterers when using air as a medium \cite{HighSpeedBEM}. 
    \par
    Predicting the scattering behaviour can be conducted by using a three step model, known as the boundary element method (BEM). Instead of the free-field propagator $\bi F_{\bi \nu \bi \tau}$, this model uses an extended propagator $\bi E_{\bi \nu \bi \tau}$, which is formed as \cite{HighSpeedBEM}
        \begin{equation}
            \bi E_{\bi \nu \bi \tau} = \bi F_{\bi \nu \bi \tau} + \bi G_{\bi \nu \bi \mu}\bi H_{\bi \mu \bi \tau},
            \label{eq:E_BEM}
        \end{equation}
    Where $\bi F_{\bi \nu \bi \tau}$ is the same as in (\ref{eq:Ftz}), $\bi H_{\bi \mu \bi \tau}$ models the propagation of the sound from the transducers to a mesh, $\bi \mu$, representing a scattering object, and $\bi G_{\bi \nu \bi \mu}$ models how the waves travel from the mesh surface to the points. Thus, the overall propagator $\bi E_{\bi \nu \bi \tau}$ considers both the direct ($\bi F_{\bi \nu \bi \tau}$) and the scattered ($\bi G_{\bi \nu \bi \mu}\bi H_{\bi \mu \bi \tau}$) components. However, while using BEM improves many applications \cite{HighSpeedBEM, ToBEMorNot}, the increased computational complexity, especially in computing $\bi H_{\bi \mu \bi \tau}$ holds back its utility \cite{ToBEMorNot}. Both $\bi G_{\bi \nu \bi \mu}$ and $\bi H_{\bi \mu \bi \tau}$ have the number of mesh elements as a dimension, and as such a mesh with many elements per wavelength (in order to improve model accuracy) will significantly increase the size of these matrices. Although, as $\bi H_{\bi \mu \bi \tau}$ depends only on the geometry of the scattering objects, it can be precomputed and cached to reduce this cost in some cases \cite{HighSpeedBEM}. However, if the object's geometry changes, $\bi H_{\bi \mu \bi \tau}$ must be recomputed, and so dynamic scenarios will suffer from the increase in computation speeds \cite{ToBEMorNot}. 
    \par
    In this paper, we will use $\bi A$ to denote an arbitrary propagator, instead of the specific $\bi F_{\bi \nu \bi \tau}$, $\bi E_{\bi \nu \bi \tau}$ or any other propagator. The index, $\bi \nu \bi \tau$, will be omitted but any propagator will be assumed to have this same form, propagating from a transducer array to a set of points. In the same way, $\bi x$ will refer to a generic set of transducer activations with the $\bi \tau$ index omitted.
    \par
   
    \textbf{Phase Retrieval Algorithms:}
        For a given propagator, the transducer activations $\bi x$, which we refer to as an acoustic hologram, need to be determined. A generalised objective function can be defined as \cite{TGS}
        \begin{equation}
            \min_\textbf{\textit{x}} \mathcal{L}(\textbf{\textit{A}}\textbf{\textit{x}}) + \mathcal{C}(\textbf{\textit{x}}).
            \label{eq:Objective_general}
        \end{equation}
        Here, $\mathcal{L}$ is an objective function of the complex field produced by the propagation of the hologram $\textbf{\textit{x}}$, for example maximising the pressure amplitude at the targets \cite{IB,GSPAT, TGS}. The constraint $\mathcal{C}$ quantifies hardware requirements for the hologram $\bi x$, often using indicator functions that are zero if these requirements are met and infinite otherwise. Common constraints include forcing transducer amplitudes to be maximum \cite{IB}, $|\textbf{\textit{x}}| = 1$, or allowing amplitudes to vary within physically feasible limits \cite{GSPAT}, $|\textbf{\textit{x}}| \le 1$. Methods to fulfil (\ref{eq:Objective_general}) take two main forms, gradient-descent as well as projective methods. 
        \par
        For a differentiable objective $\mathcal{L}$, gradient descent methods update the acoustic hologram $\textbf{\textit{x}}$ based on the computed gradients. Then, the constraint indicator function $\mathcal{C}$ can be satisfied by a projection function $\mathcal{P}_\mathcal{C}$, which projects the updated hologram into the physically feasible region defined by $\mathcal{C}$. For example, given the constraint where the amplitudes are all set to be $1$, the general projection $\mathcal{P}_\mathcal{C}$ becomes the specific $\mathcal{P}_\chi(\textbf{\textit{x}}) = \textbf{\textit{x}} / |\textbf{\textit{x}}|$, which maintains the phase of the hologram while constraining the amplitude. For a phase-only hologram and an update rate $\alpha$, this projected gradient descent update rule can be represented as
        \begin{equation}
            \textbf{\textit{x}}_{k+1} = \mathcal{P}_\chi(\textbf{\textit{x}}_k - \alpha \nabla_\textbf{\textit{x}} \mathcal{L}(\textbf{\textit{x}}_k)).
            \label{eq:PGD}
        \end{equation}
        \par
        In order to then perform this optimisation, computing the gradients can be done either explicitly or implicitly. Explicit computations require more labour from the developer \cite{AcousticElements}, where they must directly know and implement how to find the gradients. Conversely, implicit gradients are more convenient and versatile \cite{Diff-PAT, Diff-PAT-Expanded} as the exact form of the gradients are avoided by utilising automatic differentiation. This accumulates the gradients through computation of an objective and then uses the chain rule to backpropagate through this computation tree to compute the gradients of a wide variety of functions \cite{Diff-PAT, Diff-PAT-Expanded}.  This gives flexibility in algorithm design as the gradient to minimise an objective need not to be defined directly. However, to use automatic differentiation, the chain of operations needs to be maintained, requiring specific implementations such as that given in the Python packages PyTorch \cite{PyTorch} or Jax \cite{jax}.
        \par
        On the other hand, alternating projections are based on -- normally iterative -- algorithms, in which forward and backward propagations between the ultrasound source and the target points are combined with projections to feasible values. The most common algorithms are based on the Gerchberg-Saxton (GS) algorithm \cite{GS}, first described in acoustics as the iterative backpropagation algorithm (IB), which can be written as
        \begin{equation}
            \textit{\textbf{x}}_{k+1}^{IB} = \mathcal{P}_\chi(\textit{\textbf{A}}^*(\mathcal{P}_\Pi(\textit{\textbf{A}}\textit{\textbf{x}}_k | \bi y))).
            \label{eq:IB}
        \end{equation}
        In this case, $\mathcal{P}_\Pi$ is introduced as a second indicator constraint that sets the point amplitudes to some target point amplitude $\textbf{\textit{y}}$, defined as $\mathcal{P}_\Pi(\textbf{\textit{p}} | \bi y) = (\textbf{\textit{p}} / |\textbf{\textit{p}}|) \cdot \bi y$ given the point activation $\textbf{\textit{p}}$ and element wise multiplication $(\cdot)$. $\bi A$ propagates the hologram to the target points pressure, this is then constrained by $\mathcal{P}_\Pi$. This constrained point is back-propagated using the conjugate transpose of the propagator, $\bi A^*$, which is itself constrained by $\mathcal{P}_\chi$. The IB algorithm is known to act as projected gradient descent for an objective function that maximizes the sum of amplitudes at target points \cite{TGS}. 
        \par
        Despite lacking the interpretability of IB, many variations of alternating projections have been deployed for acoustic holography. For example, the multipoint Naive algorithm uses half of an IB iteration 
        \begin{equation}
            \bi x^{naive} = \mathcal{P}_\chi(\bi A^*\bi y).
            \label{eq:Naive}
        \end{equation}
        This results in very fast computation rates, but failing to finely equalize pressure at target points. Alternatively, GS-PAT removes the target points amplitude constraints, combining forward and backward propagation for high computational speed \cite{GSPAT}, while Weighted Gerchberg-Saxton (WGS) weights the pressure at each target point during each iteration to achieve low variance in pressure across target points \cite{TGS}. Nevertheless, beyond interpretability, alternating projection algorithms also lack gradient descent's adaptability to adapt to custom objectives that may be required by users for different applications.
        \par
        Alternatively, recent developments in machine learning (ML) been applied to phase-retrieval tasks for acoustic holography. For example, AcousNet which uses a convolutional neural network (CNN) in order to generate image holograms much faster than some iterative algorithms such as IB could \cite{AcousNet}. In a similar way, HU-Net \cite{HU-Net} and EIR-Net \cite{EIR-NET} both predict the transducer signals based off a target plane with EIR-Net directly improving on the performance of AcousNet \cite{EIR-NET}. ML has also been applied in the context of MHz-imaging ultrasound with the Ultrasound Deep Learning Model achieving close and well defined focal points \cite{USDL}. 
    
        \par
    \textbf{Gor'kov potential, Force, and Stiffness:} When the field has been modelled for a given set of transducer activations, various metrics can be computed that quantify the field. The simplest measure of an acoustic field is the pressure at a point in space. When the solver has focused waves at a point, there should be a local maxima of pressure at that point and the amplitude of the sound field at position can be described by $|\bi A\bi x|$. Similarly, the phase can be found by taking the complex argument of $\bi A\bi x$. 
        \par
        The Gor'kov potential is a potential field based on both the pressure and gradient of pressure surrounding a small particle as \cite{AcousticElements}
        \begin{equation}
            U = K_1 \left(|p^2|\right) - K_2 \left(\left|\frac{\partial p}{\partial x} \right|^2 + 
                                                    \left|\frac{\partial p}{\partial y} \right|^2 + 
                                                    \left|\frac{\partial p}{\partial z} \right|^2 \right),
            \label{eq:Gorkov}
        \end{equation}
        \begin{equation}
            K_1 = \frac{1}{4} V \left(\frac{1}{c_0^2 \rho_0} - \frac{1}{c_p^2 \rho_p} \right ),
            \label{eq:U K1}
        \end{equation}
        \begin{equation}
            K_2 = \frac{3}{4} V \left( \frac{\rho_0 - \rho_p}{\omega^2 \rho_0 \left(\rho_0 + 2\rho_p\right)} \right).
            \label{eq:U K2}
        \end{equation}
        Here, $c_0$ and $c_p$ are the speed of sound in the medium and the particle, $\rho_0$ and $\rho_p$ are the density of the medium and particle, $\omega$ is the angular frequency, and $V$ is the volume of the particle. $p$ is the pressure at a point and ${\partial p}/{\partial a}$ is the gradient of pressure in the $a$-axis where $a \in (x,y,z)$ \cite{AcousticElements}. The Gor'kov potential is the field such that the negative gradient of the Gor'kov potential is equal to the force acting on a small particle in the sound field. 
        \begin{equation}
            \bi F_{rad} = -\bi \nabla U.
            \label{eq:UForce}
        \end{equation}
        As a result, the point at which a particle will levitate corresponds to where the gradient of Gor'kov potential is zero (so there is no net force at that point). In reality, the levitation location will be very slightly below this point due to the force of gravity but this difference can be largely neglected for very small particles \cite{ReviewOfProgress}. The points with $\bi \nabla U=0$ includes both minima and maxima of Gor'kov potential and both are technically valid levitation locations (with net force equal to zero). However, at a maxima of Gor'kov potential, the forces diverge, and therefore the trap becomes unstable. To quantify a stable trap, we can use the stiffness at a point in space, which can be represented as the sum of the unmixed, second gradients (the Laplacian) of the Gor'kov potential as \cite{AcousticElements}
        \begin{equation}
            \nabla^2U= \frac{\partial^2 U}{\partial^2 x} + \frac{\partial^2 U}{\partial^2 y} + \frac{\partial^2 U}{\partial^2 z}.
            \label{eq:UStiffness}
        \end{equation}
        When the stiffness is at a maxima, the forces surrounding the point are strong and converging, resulting in a stable levitation trap. 

\subsection{Existing Software}
\begin{table}[]
        \centering
        \resizebox{\columnwidth}{!}{
        \begin{tabular}{|c|c||c|c|c|c|}
        \hline
        &                                                    & AcousTools & OpenMPD \cite{OpenMPD} & Ultraino \cite{Ultraino} & Levitate \cite{Levitate}     \\ \hline
        Setup & Arbitrary Boards                             & Y          & \color{red} N         & Y                       & Y                           \\ \hline
        Setup & Arbitrary Points                             & Y          & Y                     & Y                       & Y                           \\ \hline
        Propagators & Piston Model \cite{GSPAT}              & Y          & Y                     & Y                       & Y                           \\ \hline
        Propagators & BEM \cite{HighSpeedBEM}                & Y          & \color{red}N          & \color{red}N            & \color{red}N                \\ \hline
        Solvers & Naive \cite{VolDisplaySingleParticle}      & Y          & Y                     & Y                       & \color{red}N                \\ \hline
        Solvers & IB \cite{IB}                                & Y          & Y                     & \color{red}N            & \color{red}N                \\ \hline
        Solvers & GS-PAT \cite{GSPAT}                         & Y          & Y                     & \color{red}N            & \color{red}N                \\ \hline
        Solvers & Gradient Descent \cite{AcousticElements}    & Y          & \color{red} N         & \color{red}N            & Y                           \\ \hline
        Solvers & Implicit-Gradient \cite{Diff-PAT}          & Y          & \color{red} N         & \color{red}N            & \color{red} N               \\ \hline
        Analysis & Pressure                                  & Y          & Y                     & Y                       & Y                           \\ \hline
        Analysis & Gor'kov Potential \cite{AcousticElements} & Y          & Y                     & Y                       & Y                           \\ \hline
        Analysis & Force  \cite{AcousticElements}            & Y          & Y                     & Y                       & Y                           \\ \hline
        Analysis & Stiffness  \cite{AcousticElements}        & Y          & Y                     & \color{red}N            & Y                           \\ \hline
        Hardware & Control                                   & Y          & Y                     & Y                       & \color{red}N                \\ \hline
        \end{tabular}}
        \caption{Summary of various acoustic holography software libraries and packages. AcousTools meets each step of the full-stack discussed above while each other solution fails at at least part of some of the steps. AcousTools is unique in that it can be used as a complete solution that can swap and change between different propagators, solvers and contexts for various applications. }
        \label{tab:software_summary}
    \end{table}  
    Several pieces of software are available for simulating and controlling acoustic fields, and these will be compared in this section. It is expected that any additional functionality could be implemented for any piece of software, but they will be compared as presented in their original form. This is because, at least in part, the goal of a software package is to reduce the implementation burden on the user. Therefore, the comparison will not consider the potential for user-implemented extensions or the ease of doing so. 
    \subsubsection{Ultraino}
        Ultraino is an early Java-based software package that aimed to allow for start to finish acoustic development, being able to generate phases, drive transducer arrays, and simulate the resulting sound field \cite{Ultraino}. This package allows for the simulation and control of sound fields in free-field scenarios due to its simple propagation model and naive focusing method. Their framework allows for many different transducer arrays to be used, even with more obscure geometries such as a convex curved array. It also supports the computation compute some metrics, such as pressure, Gor'kov potential, and force, while lacking a stiffness measure. \par
        However, Ultraino is a stand-alone application compared to a code library, and it is not clear how a user would extend the software to support new use cases or integrate it with existing code, as no instructions are provided for doing so \cite{UltrainoGitHub}. Furthermore, there is limited support for more advanced methods used in acoustic holography, such as complex multipoint algorithms like GS-PAT or propagators that allow scattering computation like BEM. 
    \subsubsection{OpenMPD}
        OpenMPD was a later solution aiming to be a ``low-level presentation engine'' focusing on a high-speed rendering pipeline and giving access to a Unity interface for ease of use \cite{OpenMPD}. This engine supports 10kHz update rates, sufficient for persistence of vision effects \cite{GSPAT}, and is able to run lower frequency haptic content, giving them a wide variety of utility. It also provides an OpenCL implementation of the fast solver GS-PAT \cite{GSPAT} as well as code to compute all four analysis quantities that have been discussed above, from pressure to stiffness.
        \par
        OpenMPD does not allow for arbitrary arrays by default, being limited to flat 16x16 arrays. While the developers note in their paper that different configurations could be implemented, this would require changes to the OpenMPD code. As such, this does not constitute native support for arbitrary geometries. Additionally, only the piston model propagator is supplied, and thus the more complex scenarios that BEM unlocks \cite{HapticSoftness, HighSpeedBEM} are not possible with OpenMPD as provided. Similarly, only the projective solvers (i.e., GS-PAT, naive and IB) are provided, meaning any scenario that requires a different solver cannot be used.
    \subsubsection{Levitate} 
        Levitate is a python-based acoustic holography library \cite{Levitate}, which allows modelling propagation from transducer arrays to points in space based on the piston model. This library supports an explicit-gradient descent solver where the gradient must be computed by the user and does not integrate any of the more common projective methods or allow for implicit-gradient methods, which may limit its application as these are the dominant methods used across various domains \cite{HapticsMPSTM, LeviPrint, Articulev}. Levitate can compute all of the analysis metrics discussed but has no implementation for hardware support, indicating that it is strictly a software only library. Finally, it also lacks BEM support, giving it similar disadvantages discussed above. This could be a useful software package and maybe one of the most comparable to AcousTools, however, the lack of support for critical sections of the modern holography pipeline, such as many of the most common solvers, would limit its utility. 

    \subsubsection{Other Software}
    There are various other pieces of acoustic software, which mostly aim to simulate acoustic wave propagation and lack most of the features that we have identified that a developer would need. For example, K-wave is a MATLAB package that enables simulation of free-field acoustics and definition of arbitrary arrays. However, it only supports the most simple geometric solvers, does not have any scattering support and cannot compute any of the more complex analysis metrics \cite{Kwave}.  Similar issues are found in other software implementations, such as Ultrasim \cite{ultrasim} and FOCUS \cite{FOCUS}, which all lack the versatility that would be needed to give a developer the full range of options. Similarly, there are packages aim at soling BEM problems for example, OpenBEM \cite{OpenBEM}, but do not cover any other stage in the stack
     \par
    Because of the aforementioned reasons, these packages are not considered in the remaining of this work as they do not provide the user with enough of the full-stack to allow for flexible development. For a (non-exhastive) list of acoustic simulation software, which share many of these issues, see Ref. \cite{kwave-software-list}.
    \par
    Alternatively, some software acts as a single application for acoustic development instead of a library for development. One such pipeline is ArticuLev \cite{Articulev}, which allows developers to use acoustically levitated props to create display content. ArticuLev provides support for the creation of traps and props and it's methods are agnostic to the underlying algorithms that are used to generate these traps but only a multipoint version of the naive algorithm is actually provided and in a similar way, they are agnostic to the specific array but assume a 16x16 array. This is not strictly comparable to AcousTools as this is a platform built for a specific application instead of a general purpose software library but should be mentioned that for developers with specific goals, specific software may exist \cite{Articulev}.

\section{The Acoustic `Full-Stack'}
    
        Here, we present the acoustic holography full-stack, the combined process to create any given application (see Fig.\ref{fig:FullStack} Top). When creating an application that utilises acoustic holography, there are five major stages that a researcher or developer would need to go through. These are built upon one another and, of course, have some overlap due to their dependency (i.e., propagator will need setup, etc). This overlap does not invalidate the distinction between stages of the full-stack, as it is a conceptual framework to help developers understand their work instead of a hard-and-fast set of rules. 
        \par
        The acoustic full-stack is a five-step process from setting up the theoretical context (\textbf{Setup}), defining how sound propagation will be modelled (\textbf{Propagators}), solving for transducer activations (\textbf{Solvers}), analysing the existing field (\textbf{Analysis}), and physicalising this on real device (\textbf{Hardware}). As the developer progresses in their application, their work turns from more conceptual to more physical, the acoustic full-stack also follows this trend.
        
    \subsubsection{Setup}
        The bottom of the stack is the setup phase, defining the transducer array's shape and parameters as well as the locations in space where acoustic waves are focused \cite{LeviPrint, AcousticElements}. 
        \par
        The transducer arrays used across acoustic applications vary from flat square boards \cite{OpenMPD} to rectangular \cite{AUTD3} and curved surfaces \cite{TinyLev}. Even more complex geometries have also been investigated in theory, such as spherical or triangular arrays, but these are rarely used in practice due to the added complexity \cite{LeviPrint}. Because of the variety of different options, the first decision that a developer must make is which of these arrays to use, and the software they use should accommodate this. 
        \par
        After defining the array, still as part of setup, the developer needs to conceptually define where in space the acoustic waves should be focused. The area a given transducer array can focus will vary based on its geometry. For example, a single-axis device \cite{TinyLev} has significantly less freedom in focusing locations compared to a larger 3D array \cite{OpenMPD, AUTD3}. Therefore, the transducer geometry needs to be accounted for when defining where the focal points can be created.  
        \par
        Finally, the various acoustic parameters and constants need to be defined, such as the frequency of sound; most often 40kHz \cite{OpenMPD, Ultraino, AUTD3, GSPAT, VolDisplaySingleParticle, HapticsMPSTM, LeviPrint} in holography but other domains use other frequencies. Medical applications often use much higher frequencies in the MHz range \cite{ZebrafishBubbles, LivingLevitation, BEMImage} and holography using a water medium can use lower frequencies, potentially even audible waves \cite{MomentumShaping}. Additionally, the reference pressure of transducers at 1m \cite{GSPAT}, the density of the medium and levitated objects, and the speed of sound in that medium \cite{BruusAcoustofluidics} are all parameters to be set. These are often not choices a developer would actively make, but rather physical properties of the system they have selected, for example the frequency of waves used are dependant on the resonant frequency of the transducers. However, these parameters should be acknowledged, as earlier design decisions will influence them. 
       
    \subsubsection{Propagators}
        Once the transducers and points have been defined, the next step is to couple them together by modelling how acoustic waves travel from transducers to points \cite{GSPAT, Diff-PAT, HapticsMPSTM, HighSpeedBEM, ToBEMorNot, HapticSoftness}. There are two main scenarios that can be used, free-field with no scattering objects and propagation with obstacles acting as scatterers. The free-field case, using the piston model as described in (\ref{eq:Ftz}), is fairly easy to compute and can be calculated at very high speeds for real-time dynamic holography \cite{GSPAT}. It is versatile and, in most free-field cases, the scattering effects can be neglected as they do not impact performance \cite{GSPAT, LeviPrint}. 
        \par
        However, in some cases, scattering effects are too significant to ignore \cite{HighSpeedBEM, ToBEMorNot}. Thus, the developer may choose to use a propagator that models this scattering. Given the model of the scatterer as a mesh, BEM can be used to predict how the sound scatterers off the boundary of the object \cite{HighSpeedBEM} as seen in (\ref{eq:E_BEM}). Accounting for them can bring improvements in a multitude of applications \cite{HapticSoftness, ToBEMorNot, HighSpeedBEM} but come at a significant computational cost \cite{ToBEMorNot}.
        \par
            \begin{figure}
                \centering
                \includegraphics[width=0.9\linewidth]{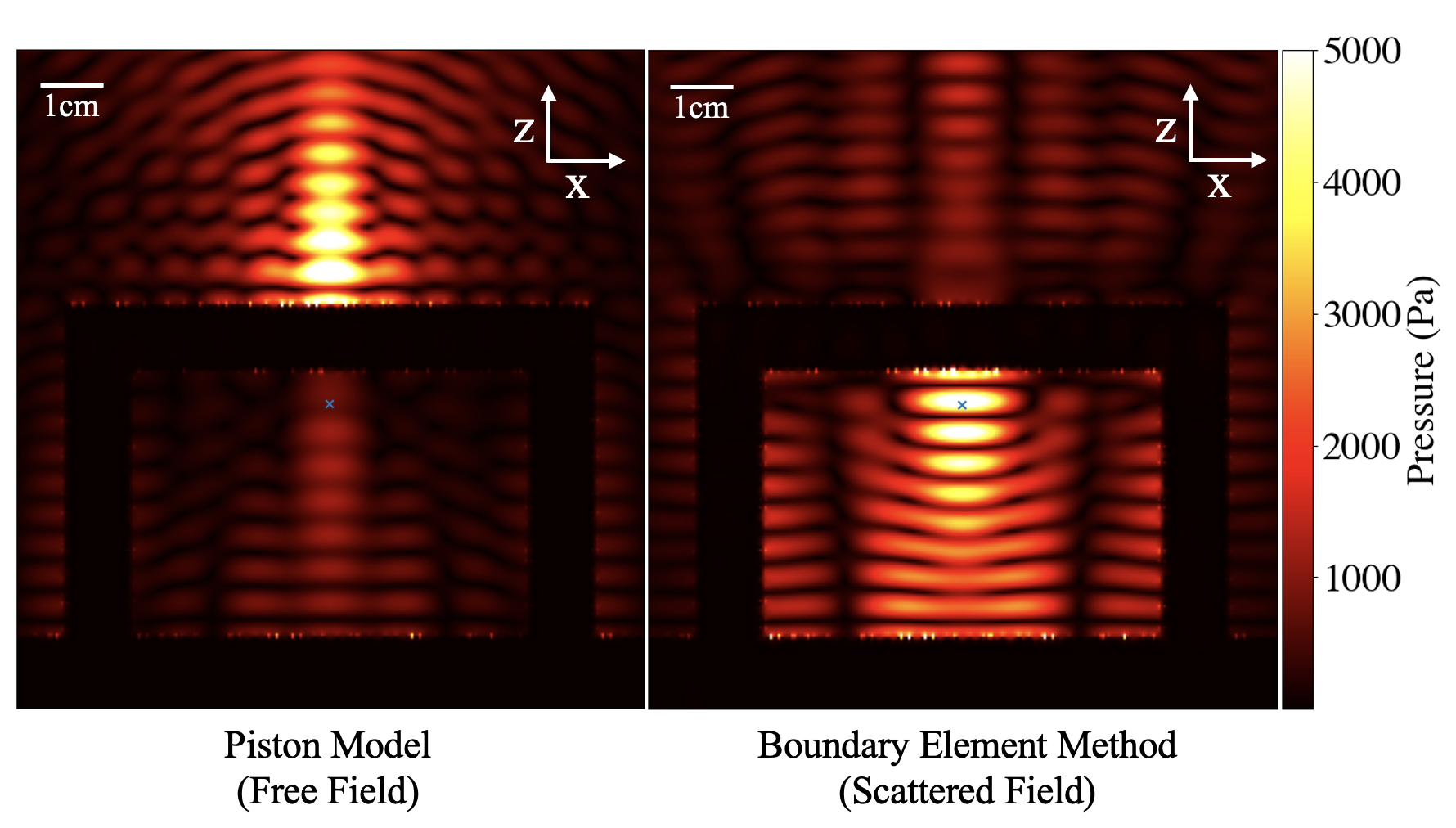}
                \caption{Comparison of using the free-field, piston model (left) and the boundary element method (right), when generating a focal point below a sound-scattering overhang (the target position is marked with a blue cross). Both are rendered using BEM to show the discrepancy caused from solving for transducer activations with an insufficiently accurate model. In the piston model case, the waves are reflected off of the top of the bridge section as the propagator did not consider it, meaning waves cannot be focused at the target. Contrastingly, BEM is able to take into account the reflections and focus at the desired location. Note both images share a single colour bar}
                \label{fig:BEMvsPM}
            \end{figure}
        \par
        The selection of propagator is therefore a defining element of the process of creating an acoustic application. Using an inappropriate method can lead to large inaccuracies in scattered field as shown in Fig.\ref{fig:BEMvsPM}, where the piston model case gives a significantly different and inaccurate result compared to the BEM propagator. Having multiple options that can easily be swapped and compared enables a developer to weigh the benefits and select the one that is best for their context. Being forced to use a single propagation method (normally by being limited to the free-field case) will restrict where their application can be used.  
        
    \subsubsection{Solvers}
        Once the setup and propagator have been defined, acoustic solvers compute the transducer activations that cause transducers to focus sound at the chosen points \cite{VolDisplaySingleParticle, GSPAT, ToBEMorNot}. Supporting a number of solvers enables a developer to have different options, but determining the best solver requires careful analysis of the problem the developer is trying to solve. For an application that requires fast computation speeds, an algorithm such as GS-PAT \cite{GSPAT} may be best. On the other hand, WGS \cite{TGS} gives high pressure and low variance across multiple points. For more complex scenarios, a implicit-gradient descent optimiser can solve more complex objectives \cite{Diff-PAT}. It is important for a developer to be able to define custom objective functions for their own context and the developer must weight which is best, potentially comparing multiple options before a single best choice is decided upon. This emphasises the desirability for simple swapping between solvers, to aid in this development process.

    \subsubsection{Analysis}
        At this point, a developer has defined their setup, propagators, and solvers to generate acoustic focal points at the desired locations. After this, they may need to quantify certain properties of the sound field (e.g., pressure, Gor'kov potential, force, stiffness), either for validation or further optimisation \cite{AcousticElements, AcoustoFab, HighSpeedBEM, LeviPrint}. Importantly, this step may overlap with the Solvers step. For example, using a custom objective defined to maximise the pressure at a point may require elements from both the previous and current layer. However, this layer can also sit by itself, for example, quantifying the force on a particle to validate if a particle can be correctly levitated at the desired location \cite{Ultraino}. The quantities discussed here, which help to quantify an acoustic field, are pressure as well as Gor'kov potential, force, and stiffness as described in (\ref{eq:Gorkov}), (\ref{eq:UForce}), and (\ref{eq:UStiffness}) respectively.
        \par
        
        \par
        Which one of these metrics is the most useful will depend on the use case a specific developer might have. If they are developing a haptic interface, the pressure at a point might be the only thing that they need to consider \cite{HapticsMPSTM} while for levitation the Gor'kov potential, force, and stiffness might become more appropriate \cite{HighSpeedBEM,AcousticElements}. Additionally, the square of pressure correlates very well with stiffness, and as such, the simpler metric can be used as a proxy for the more complex one in practice \cite{GSPAT}. As discussed above, having a diverse set of tools means that the correct tool can be used at the right moment.
        
    \subsubsection{Hardware}
        The final step before an acoustic holography application can be deployed is to control the physical transducers in the real world. This involves sending synchronised phase and amplitude values from a computer to the device, physically realising the until-this-point simulated fields. This presents difficulties as for the very high update rates that acoustic holography often demands, up to 10kHz for levitation, a high degree of synchronization is needed. Various platforms exist to address this problem. Such platforms include OpenMPD \cite{OpenMPD}, which can run at 10kHz; AUTD3 \cite{AUTD3}, which is able to run at 1kHz; and Ultraino \cite{Ultraino}, which runs at 25kHz. These different platforms also have different ways to communicate between the transducer arrays and the computer, which would need to be accommodated by any software used. 
        \par
        This section is more difficult for software packages to be general purpose, as each different hardware setup has a different communication method and structure for the messages to be sent \cite{OpenMPD, AUTD3}. Thus, many software packages only function for one or a small number of hardware devices but having at least one supported hardware device enables the simulated fields to be physicalised in real world.
   
\section{AcousTools}
\begin{figure*}
        \centering
        \includegraphics[width=\linewidth]{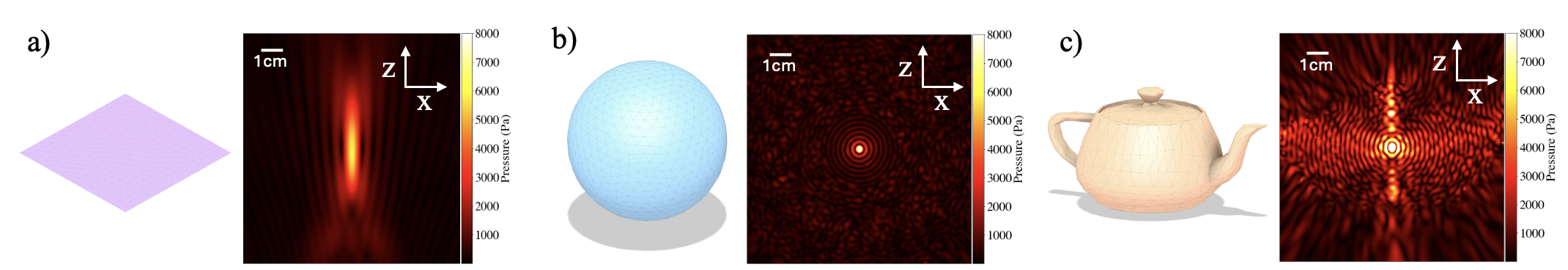}
        \caption{Examples of using different meshes as the definition for transducer arrays creating a focal point in the origin of the working area (high pressure region at the centre of each subfigure). For a given mesh, a transducer is placed at each cell centre with the normal of the mesh defining the transducer normal, with normals pointing into the working volume inside the mesh. a) flat array similar to a common setup used in practice \cite{AcousticElements}, b) Sphere c) Teapot. For each array, the mesh is scaled so that the limits of the mesh in the x-axis is $\pm$11.8cm, centred at the origin (except the flat array which is translated down in order to give a working volume above it) and the normals point towards the origin. }
        \label{fig:MeshBoardsFig}
    \end{figure*}
As discussed above, the existing software implementations for acoustic holography are limited in their functionality, often not supporting a full range of solvers or propagators. To address this, we present AcousTools; a python library built to meet the full-stack of acoustic holography. AcousTools is open source and can be found at our GitHub repository \cite{AcousTools}.
\par
Python was chosen to implement AcousTools because of its reputation for accessibility, especially for novice users \cite{pythonEase}. It is one of the most widely used programming languages \cite{PythonUse} and among the fastest for beginners to learn. Whereas C++, commonly used in acoustic holography, is considered one of the most difficult \cite{PythonLearning}. This popularity, combined with Python's ease of adoption and of use, suggests that adoption and extension of AcousTools would be similarly straightforward. 
\par
Additionally, AcousTools is based on PyTorch \cite{PyTorch} due to its parallelised Tensor implementations, GPU support, and automatic differentiation system. These together make PyTorch an ideal base for AcousTools and provide the secondary benefit that potential future machine leaning applications for acoustic holography could be easily implemented, as PyTorch is widely used in this domain. 
\par
Here we discuss AcousTools' ability to meet every step of every stage of the acoustic full-stack discussed above. Table \ref{tab:software_summary} summarises both the existing software's performance and AcousTools' features, illustrating that AcousTools meet every stage while the others all lack in some area or another. There is also a lack of standardisation across acoustic holography research \cite{OpenMPD, AUTD3, Ultraino}, and we believe that AcousTools with its general purpose nature could fill this gap for researchers across all domains.

    \subsection{Setup}
    AcousTools can facilitate the setup-phase of the full-stack in each of the ways discussed above. Custom transducer arrays can be defined using a mesh, interpreting the data as a collection of points with their corresponding normals. Using this, a transducer can be placed at the centre of each of the mesh elements with the normals used as the transducer normals. Various examples of this can be seen in Fig. \ref{fig:MeshBoardsFig}, with arrays ranging from a simple flat board to more complex shapes like a teapot. In AcousTools, the code for this is simply

        \begin{verbatim}
sphere_board, sphere_norms 
    = mesh_to_board(
        './data/Sphere-lam1.stl')
        \end{verbatim}
    This code loads the mesh and collects the positions and normals. In addition to custom transducer arrays, Acoustools has three common transducer configurations pre-configured, all based on arrays of 256 transducers arranged into 16x16 grids. This includes a single array positioned below the working volume and directed up \cite{AcousticElements}, a single array positioned above the working volume and pointed down \cite{HighSpeedBEM}, and one with the combination of these two arrays pointing towards one another \cite{GSPAT}. These are supported as pre-constructed variables without needing to load a mesh, for example:
\begin{verbatim}
board = TOP_BOARD
\end{verbatim}
    \par
    After the boards are defined, points at which sound should be focused. To define a single point, $N=1$, located at $\bi z = (x_t, y_t, z_t)$ can be simply defined as
\begin{verbatim}
p = create_points(N=1,x=xt,y=yt,z=zt)
\end{verbatim}%
    In a similar fashion, multiple points can be configured, passing the same function larger values for $N$.
    AcousTools also has the ability to be configured to work with different transducer frequencies which are used across different applications. The wavenumber, $k = 2\pi f/c_0$, can be used to specify the resonant frequency of the transducers being used - most often 40kHz for holography in air \cite{GSPAT, AUTD3, Diff-PAT, OpenMPD, AcousticElements} but different domains use different frequencies such as the MHz range used in medical applications \cite{BEMImage, ZebrafishBubbles, LivingLevitation}. Given some wavenumber, $k$, AcousTools can solve for a hologram at that frequency simply using
\begin{verbatim}
x = wgs(p, k=k)
\end{verbatim}
    AcousTools is able to operate at a range of frequencies, even into the MHz range meaning it could be used to develop applications across a number of domains. In theory, any frequency can be used but for different frequencies, using different transducers, the properties of the device may change and therefore models such as the piston model in (\ref{eq:Ftz}) may need to be swapped for a different model. Additionally, different frequencies will produce different patterns with higher frequencies naturally having smaller wavelengths and therefore finer-resolution focusing patterns (assuming equal arrays). This means that for applications that care about the size of a focal point, testing different frequencies maybe advantageous but comes with downsides. For example, a finer focus (with higher frequency waves) may increase resolution at the cost of increasingly co-linear waves reducing the ability to focus across a wide area due to the tightening of the piston source (\ref{eq:Ftz}). When changing the frequency it is likely a whole new array would be needed and as such its parameters would also change. The separation of the transducers, their size and shape or the power they output all would need to be adjusted. These parameters all can be provided to AcousTools to model the behaviour of these arrays.
\subsection{Propagators} 
        As discussed above, the two main propagators used for acoustic holography are the piston model and the boundary element method. The piston model, as described in  (\ref{eq:Ftz}), uses a free-space model and is implemented in AcousTools as 
\begin{verbatim}
F = forward_model_batched(p,board) 
\end{verbatim}
        As shown in Fig. \ref{fig:BEMvsPM}, the piston model is too simple to accurately model sound propagation in various situations. In these cases, BEM can be used to compute a propagator that captures sound scattering, using a mesh to model a scattering object. 
\begin{verbatim}
scatterer = load_scatterer(
            "./data/Sphere-lam2.stl")
E = compute_E(scatterer, p, board)
\end{verbatim}
    Because these propagators share the same structure - both are matrices where the complex pressure at a point is obtained by the matrix-vector product of the propagator and the transducer activations - these can be swapped without requiring additional changes to the rest of the code. This allows a user to compare the two contexts with minimal changes to their code. For some solver (e.g., IB \cite{IB}), the propagator is the only component that needs to be interchanged between the piston model and BEM contexts.
\begin{verbatim} 
x = iterative_backpropagation(p, 
                    board=board, A=F)
x = iterative_backpropagation(p, 
                    board=board, A=E)
\end{verbatim}
\par
These propagators all assume that the medium is homogeneous - for other applications that this assumption is not true, such as in-vivo applications, other applications may need more complex propagators. Critically though, if they take the same form as the existing models, they can then be used with the rest of AcousTools as with any other propagation matrix.  
    \subsection{Solvers}
    AcousTools provides a number of solvers, giving further flexibility depending on the context of an application. High-speed solvers, such as multipoint naive and GS-PAT, offer quick computation \cite{GSPAT}, while slower solvers, such as WGS, can generate focal points with small variance of pressure \cite{TGS}. As discussed, the choice of which solver to use is a critical stage in the development process, and a wide variety that can be easily compared enables a developer to make an informed choice. 
    \par
    Using AcousTools, a developer can simply swap the solvers for one another to compare their performances as
\begin{verbatim}
x_ib = iterative_backpropagation(p, 
                    board=board, A=A)
x_gspat = gspat(p, board=board, A=A)
x_wgs = wgs(p, board=board, A=A)
\end{verbatim}
    These lines compute the transducer activations for a given board and propagator (\verb|A|) with IB, GS-PAT, and WGS, respectively. These can be swapped in and out as the developer desires, making testing and iteration easy.
    \par
    AcousTools can leverage implicit-gradient descent algorithms to optimise more complex objectives \cite{Diff-PAT, Diff-PAT-Expanded}, which can go beyond what projective algorithms can do. Implicit gradient-descent methods can minimise Gor'kov potential directly, produce fields with specific metrics at specific points, and even combine multiple objectives simultaneously. Although these methods are often slower than others \cite{Diff-PAT} and may not be suitable for all cases, their ability to use custom objectives provides greater flexibility for the developer. 
    \par
\begin{figure}
    \centering
    \includegraphics[width=\linewidth]{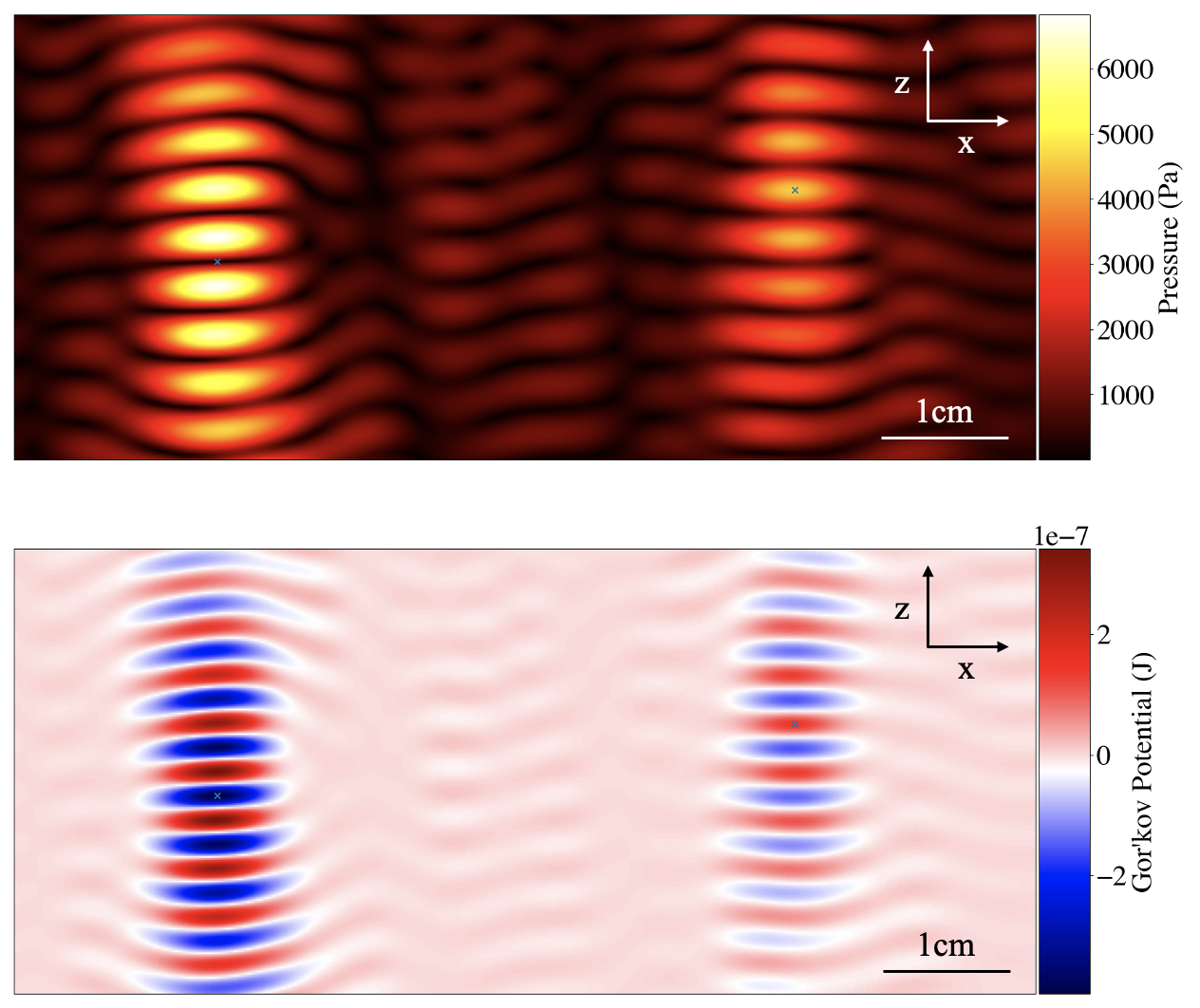}
    \caption{The result of using the gradient descent solver to minimise Gor'kov potential at one point and maximise pressure at another. This can be seen in both the top (pressure) figure, with a area of low pressure surrounded by high pressure lobes at one point while the other has high pressure at the target. The bottom figure (Gor'kov potential) also shows this with a minima of Gor'kov potential at one point and a maxima at the other.}
    \label{fig:GradSolver}
\end{figure}

    To utilise the implicit-gradient solver, a developer must first define an objective, for example, to minimise Gor'kov potential at one point while maximising pressure at another. The objective function for this could be expressed as 
    \begin{equation}
        \mathcal{L}(\textit{\textbf{x}}) = -|\textit{\textbf{F}}_f\textit{\textbf{x}}| + \lambda U(\textit{\textbf{F}}_t\textit{\textbf{x}}).
        \label{eq:trap_focalpoint}
    \end{equation}      
    Where $\textit{\textbf{F}}_f$ propagates to the target focal point and $\textit{\textbf{F}}_t$ propagates to the target trap. This objective can be written in AcousTools as
    \begin{verbatim}
    
def objective(transducer_phases, points, 
            board, targets = None, 
            **objective_params):
    lambda_constant = 
                objective_params['lambda']
    p1 = points[:,:,0].unsqueeze(2)
    p2 = points[:,:,1].unsqueeze(2)
    pressure = propagate_abs(
                transducer_phases, 
                p1, board=board
                ).squeeze()
    U = gorkov(transducer_phases, p2, 
                board=board ).squeeze()
    return (-1*pressure+
        lambda_constant*U).unsqueeze(0)
\end{verbatim}
    Then this can be sent to the implicit-gradient descent solver to be optimised, the exact optimiser and parameters can be customised such as the optimiser, the learning rate, weights of each term in the objective, and learning rate schedulers. These parameters can then be passed to the solver:
\begin{verbatim}
p2 = create_points(2, y=0, 
        max_pos=0.04, min_pos=-0.04)
x = gradient_descent_solver(p2, 
        objective=objective, board=board, 
        objective_params={
            'lambda':1.32e10}, 
        iters=1000)
\end{verbatim}  
This results in the sound field shown in Fig. \ref{fig:GradSolver}, using the Adam optimiser \cite{Adam}, a learning rate of $0.01$, $1000$ iterations and $\lambda = 1.32\times10^{10}$ which ensures each component has a similar magnitude. One point has a minima in the Gor'kov potential corresponding to a low pressure region sandwiched between two high pressure regions and the other point has a maxima of Gor'kov potential corresponding to a region of high pressure, as defined in the objective. This could be useful for multi-modal applications where a focal point is used for haptic applications while the minima of Gor'kov potential corresponds to a levitation trap that can hold a small bead \cite{AcousticElements}. 
\par
Projective solvers, such as IBP and GS-PAT, could not fully create the field in this situation, as they produce high pressure regions at their targets. These can be converted to traps through levitation signatures \cite{AcousticElements}, but such signatures convert all focal points instead of only one. As seen in Fig. \ref{fig:GradSolver}, a focal point has a local minima just below the high pressure regions, and so a similar result to Fig. \ref{fig:GradSolver} could be produced by focusing slightly above the target trap position. However, using this method may lead to both sub-optimal solutions with smaller forces and is harder to customise the relative strengths of the trap and focus. Therefore, the implicit-gradient solver is an additional tool for a developer, and combining this with the other methods provides a full toolbox to use.
    \par
    The main other category of phase-retrieval algorithms is ML based methods. As discussed, these use large networks to predict the phases that create some target image in the focal plane. These are often specialist networks and are not directly implemented in AcousTools but because it is built upon PyTorch, every element of AcousTools is able to be used with the automatic differentiation engine, meaning ML components can be designed, trained and deployed using it. Because of this, AcousTools can facilitate the growth of this corner of the acoustics world - providing a combined and unified platform for research.
    \subsection{Analysis}
\begin{figure}
            \centering
            \includegraphics[width=\linewidth]{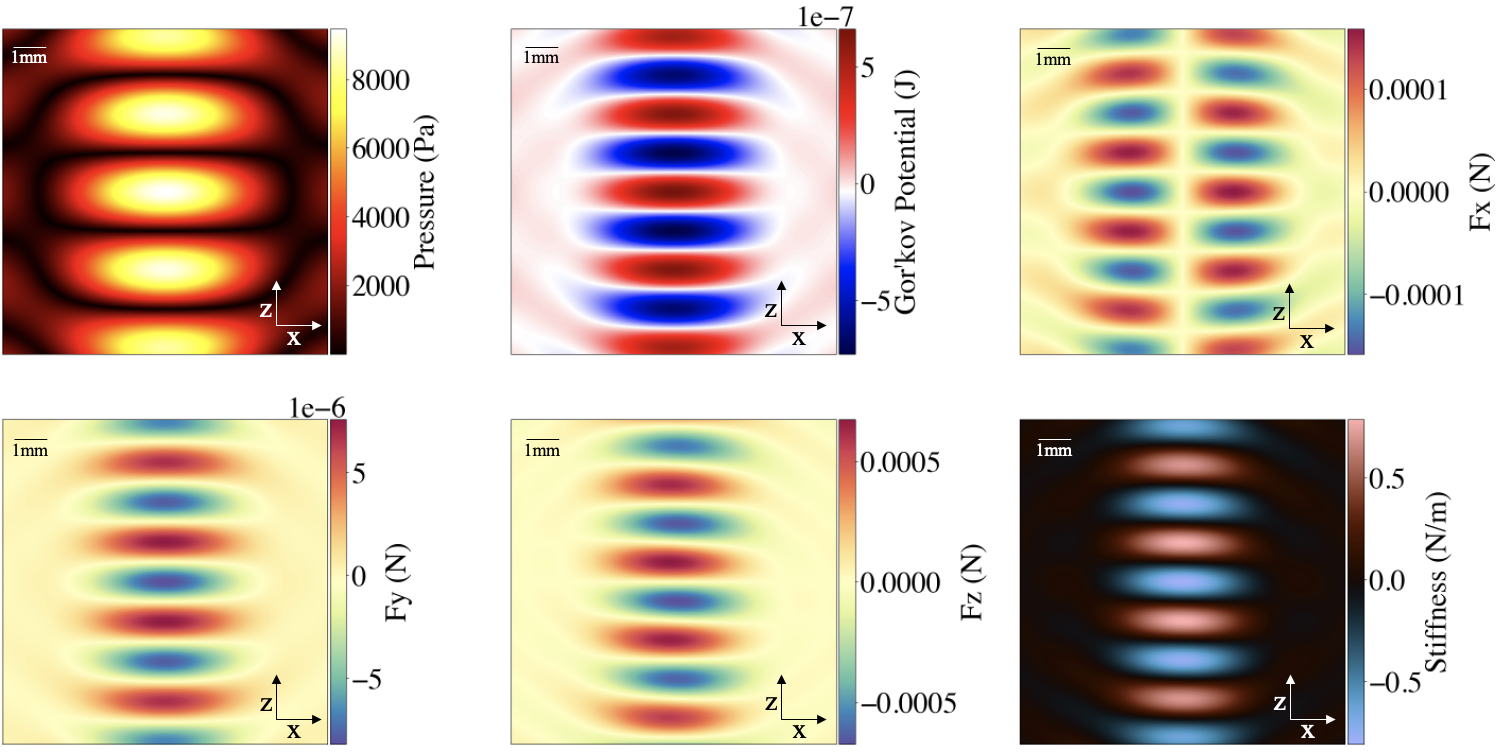}
            \caption{The various analysis metrics that have been discussed visualised in the xz plane (constant y=0). A minima of pressure surrounded by high pressure corresponds to a minima of Gor'kov potential which denotes areas of zero net-force (neglecting gravity). The regions that a particle would be trapped for levitation also correspond to the areas where force converges (positive displacement leads to a negative force and visa-versa) which leads to a high value for stiffness. Note, the visualisation plane is offset from the plane the focal point sits in to avoid very small forces.}
            \label{fig:Analysis}
        \end{figure}

        Many of the analysis methods have already been demonstrated here, pressure amplitude (see Figs. \ref{fig:BEMvsPM}, \ref{fig:MeshBoardsFig}, and \ref{fig:GradSolver}) and Gor'kov potential (see Fig. \ref{fig:GradSolver}). The pressure amplitude and Gor'kov potential can be simply computed as
        \begin{verbatim}
pressure = propagate_abs(x, p, board=board)
U = gorkov(x, p, board=board)
        \end{verbatim}
        Alternatively, the complex pressure that maintains phase information can be computed as
        \begin{verbatim}
pressure = propagate(x, p, board=board)          
        \end{verbatim}
        Computing Gor'kov potential here uses analytical gradients of the piston model instead of using finite differences and makes it significantly faster compared to the standard method using finite differences \cite{AcousticElements, HighSpeedBEM}. The speed-up between the two methods is significant, the analytical method is able to compute 4232 solutions per second compared to 1922 when using finite differences; these values were computed over 3000 random independent points, computed sequentially. A similar analytical solution is provided using the analytical gradient of BEM. For both propagators, an analytical method for computing the gradient of Gor'kov potential is also provided for computing force. A summary of these metrics can be seen in Fig. \ref{fig:Analysis}, showing that all these metrics from pressure to stiffness. In this case, the propagator used is the piston model and metrics use the analytical gradients where possible, similar visulisations could be created using BEM by simply swapping the propagators as discussed above. 
        \par
        For a developer, the ability to select the appropriate metric for a given situation ensures that the acoustic behaviour can either be quantified after the fact to verify correct performance \cite{Ultraino} or included in the solver to allow explicit optimisation of these quantities \cite{VolDisplaySingleParticle, Diff-PAT, AcousticElements}. Once again, a versatile selection leads to more versatile applications for developers.
    \subsection{Hardware}
    The final stage of the full-stack is Hardware, sending the transducer activations to a real world device. As discussed, this is difficult for a software package to be fully general purpose as every device may have a different communication protocol with little standardisation. Therefore, AcousTools uses OpenMPD's board drivers \cite{OpenMPD} to drive the 16x16 transducer arrays used in that paper. This is because these arrays are common across a number of applications, from levitation \cite{VolDisplaySingleParticle, GSPAT}, haptics \cite{ HapticsMPSTM, ToBEMorNot} to fabrication \cite{AcoustoFab}, have been the target of many phase retrieval methods \cite{Diff-PAT, GSPAT, VolDisplaySingleParticle} and propagation methods \cite{AcousticElements, HighSpeedBEM}, and are based on an open source platform \cite{Ultraino, OpenMPD}. Sending the activations is simple with AcousTools. All the developer must do is connect to the device and send the computed activations:
\begin{verbatim}
lev = LevitatorController()
p = create_points(1,x=0,y=0,z=0)
x = gspat(p)
lev.levitate(x)
input() #Wait for user input to disconnect
lev.disconnect()
    \end{verbatim}
    \begin{figure}
        \centering
        \includegraphics[width=\linewidth]{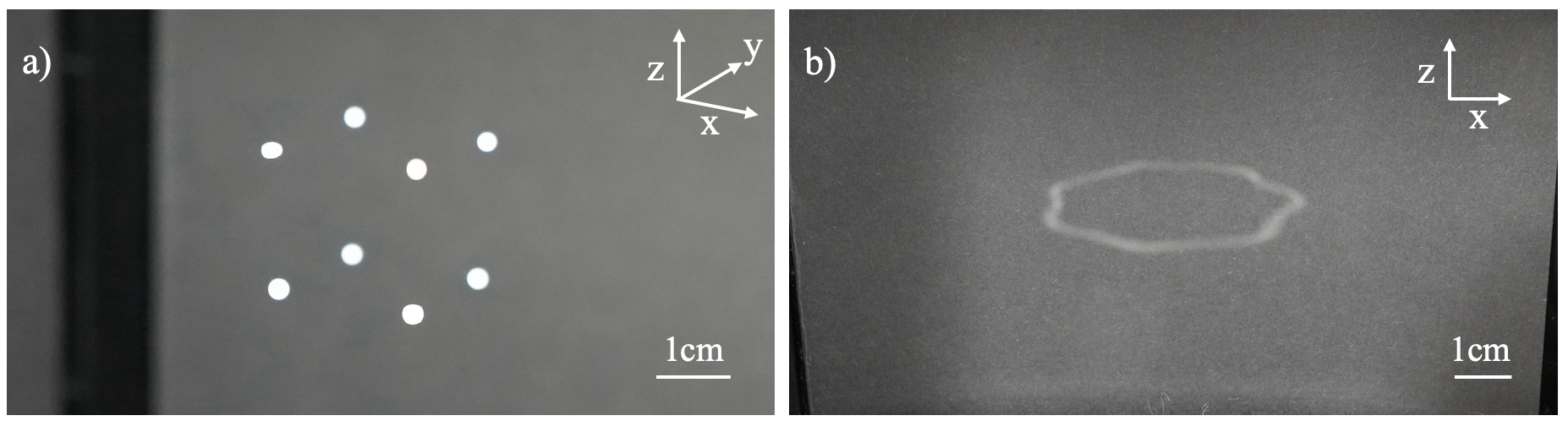}
        \caption{Examples of real-world levitation prodiced with AcousTools using two opposing transducer arrays. a) 8 points forming a static cube. b) A circle rendered using one particle moving at high speed, using persistence of vision effects to produce the illusion of a single shape.}
        \label{fig:LevitationExamples}
    \end{figure}
    Fig. \ref{fig:LevitationExamples} shows examples of levitation produced by AcousTools, with multipoint free field levitation (Fig. \ref{fig:LevitationExamples}a) and high-speed volumetric content (Fig. \ref{fig:LevitationExamples}b). In addition, BEM-based levitation is shown in Fig. \ref{fig:bunny_force}, and Fig. \ref{fig:DropletSim} shows the versatility of domains AcousTools can be used for.

    \subsection{Application: Droplet-Solid Levitation}
\begin{figure}
            \centering
            \includegraphics[width=\linewidth]{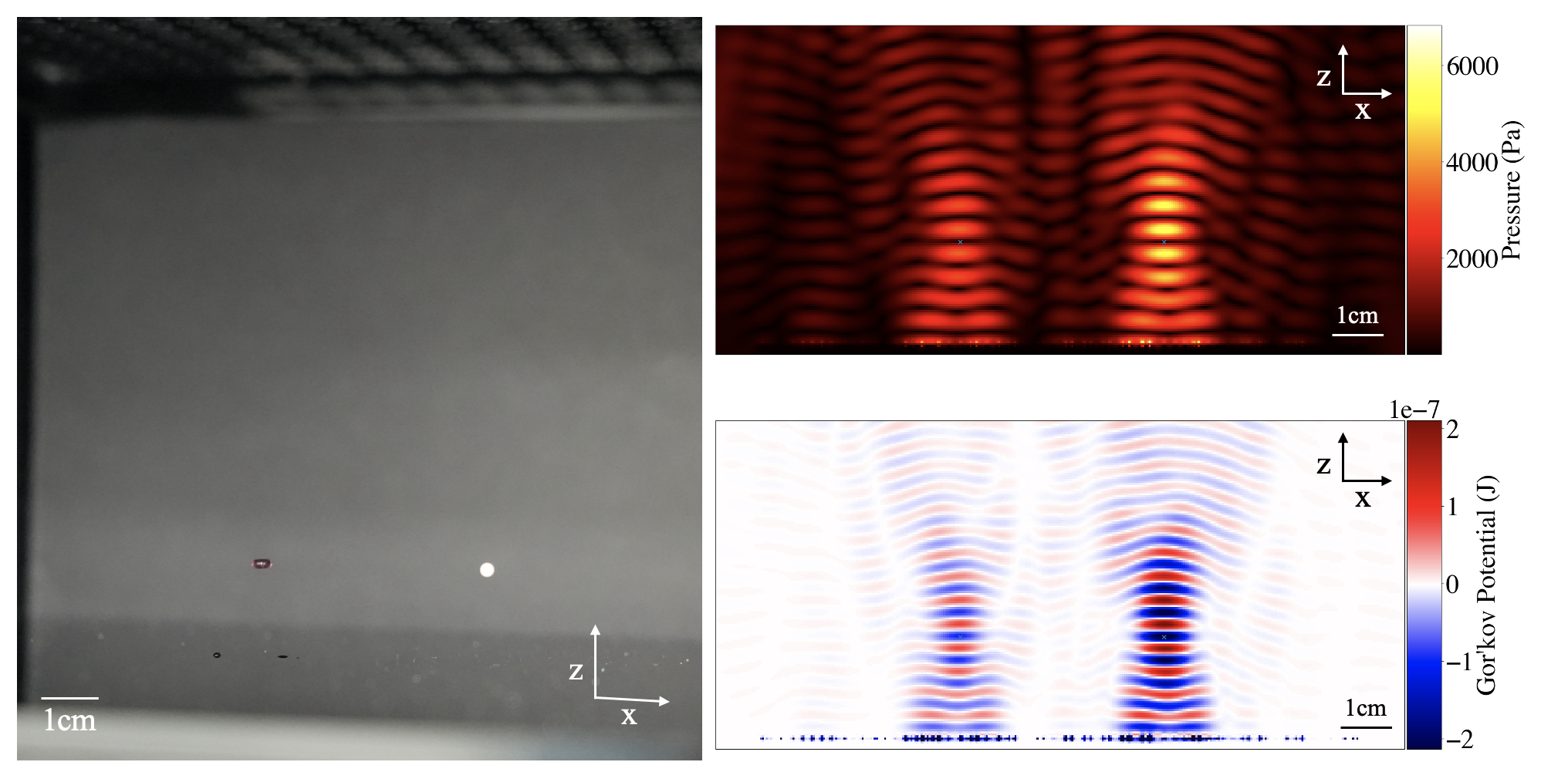}
            \caption{The results of optimising two traps, one for droplet levitation and one for solid particles. The droplet trap has $U = -9.98\times10^{-8}J$ and the solid particle trap has $U = -2.5\times10^{-7}J$, indicating that the droplet trap will apply less force and will therefore not burst the droplet. The real-world levitated results are shown (left) as well as the simulated pressure (top-right) and Gor'kov potential (bottom-right) fields}
            \label{fig:DropletSim}
        \end{figure}

        To demonstrate AcousTool's full-stack potential, we optimise two traps above a flat reflector with a downwards-facing transducer array. These traps are optimised so that one levitates a solid particle and the other levitates a droplet of liquid, with the final results seen in Fig. \ref{fig:DropletSim}. 
        \par
        The acoustic levitation of droplets is desirable from applications from fabrication \cite{Levitate, AcoustoFab} to mid-air chemistry \cite{AcousticChemistry} but is more challenging than solid particle levitation because too strong of a force applied to the droplet may cause it to buckle and burst \cite{Droplets, HighSpeedBEM}. This can be mitigated by optimising the trap to have a reduced Gor'kov potential applying a controlled force to the droplet. On the other hand, the solid particle is best trapped by minimising Gor'kov potential and therefore applying high, converging forces. If traps of the same strength are used for both traps, the droplet would be at risk of bursting if both are strong or the solid bead would have sub-optimal forces if the traps are weaker, meaning if the particle is moved then it will be at risk of being dropped \cite{VolDisplaySingleParticle}. Careful control of the forces gives a best-of-both-worlds situation, with the forces being as strong as possible for both traps without harming reliability \cite{Droplets}.
        \par
        \par
        As AcousTools can achieve all the stages of the full-stack -- from defining the context up to physicalising the levitation -- droplet levitation can be easily deployed using an implicit-gradient descent solver. By defining an objective that computes the Gor'kov potential at two points, two different traps can be created. This objective seeks to minimise the potential at one point while it minimises the square error between the potential at the other and a given target value. Using an appropriate coupling parameter produces two traps, one strong for the solid particle and the other with a weaker trapping strength for the droplet. When rendered in reality, the droplet stays intact, and both the particle and droplet can be rendered together. The objective for this can be described mathematically as:
        \begin{equation}
        \mathcal{L}(\textit{\textbf{x}}) = U(\textit{\textbf{E}}_1\textit{\textbf{x}}) + \lambda  || U(\textit{\textbf{E}}_2\textit{\textbf{x}}) - U_{target}||_2^2.
        \label{eq:dropletLev}
        \end{equation}          
        Here, $\bi E_1$ and $\bi E_2$ are the BEM propagators that model the scattering from a flat reflector to each of the target points. The resulting levitation as well as the simulated pressure and Gor'kov potential are shown in Fig. \ref{fig:DropletSim}. This was conducted using a $U_{target} = -1\times10^{-7}J$ for the droplet trap and $\lambda = 7 \times10^8$. A learning rate of $1$ was used and the optimisation was performed over $1000$ steps with the Adam optimiser \cite{Adam}. The droplet trap has a resulting Gor'kov potential, $U = -9.98\times10^{-8}J$, $99.8\%$ of the desired target.
        \par
        This would not be possibly using OpenMPD, Ultraino or the other ultrasonic software packages that have been discussed. AcousTools is the only option for a developer wanting to develop applications that exist outside of the limits of what has previously been common. Instead of being limited to just free-field modelling with solid particles, AcousTools is truly versatile, giving a developer a wide variety of tools for their context. 
        
\section{Discussion}
The acoustic full-stack gives a framework that researchers can use to evaluate their acoustic applications. When developing a novel method for some layer in the stack, it can be used to narrow down on comparable existing solutions. This framework will give clarity to field of acoustic holography and provide a lens through which to view their work.
\par
Existing software packages to give developers to access acoustic holography historically did not meet all the criteria in the stack. As seen, they often fail to provide sufficient options, especially in the Propagators and Solvers stages (e.g., BEM support is the most neglected feature). In contrast to this, AcousTools specifically allows for full-stack development -- providing options to a developer. They can use the appropriate methods for their application instead of simply being limited to what is available in the software they are using. Even if the user could implement it themselves, one of the goals of using existing software is to reduce load on the user and as such software not providing these tools is a significant downside - one which AcousTools avoids. 
\par
Developers from a wide variety of fields can take AcousTools and deploy it in their contexts. Python as a language is among the easiest languages to learn \cite{pythonEase} and is one of the most used languages \cite{PythonUse} meaning there is a very large number of people who would be able to take AcousTools and integrate acoustic holography into their own domain. A haptician could compare the effects of different transducer board or different propagators on immersion, someone interested in levitation could optimise the shapes that they render to give higher visual quality or a chemist could control small amounts of liquid drops for non-contact mid-air reactions while maintaining specific forces. AcousTools' use of PyTorch not only provides the ability to integrate implicit-gradient solvers as discussed above but also allows users to deploy a machine learning solutions to their work. Novel solutions that have already been shown to be helpful for various acoustic domains \cite{AcousNet, StableLev}. Thanks to AcousTools, possible applications in acoustic holography are not limited by existing software and now depend on developers alone.

\par
We hope that AcousTools will become the standardised code library for acoustic holography for all applications. Using consistent code across researchers and industry developers would facilitate greater interoperability and reproducibility, which is currently hindered by the fact that users mostly use their individual implementations. Of all existing software, AcousTools is the clear choice for this standard due to the aforementioned versatility and accessibility. As a greater degree of standardisation occurs, more work can be put into AcousTools by the whole community, extending the features it supports and making it even more suitable to be the standard software library.

\section{Conclusion}
In this paper, we introduced a framework, which we refer to as the acoustic full-stack, for acoustic holography applications. This framework guides developers through the stages required to transform an idea into a tangible, real-world device. The full-stack provides a structured approach for comparing software packages, and our analysis shows that no existing software allows developers to make informed decisions at every stage of the process. To address this gap, we presented AcousTools, a python-based acoustic holography library that allows developers to explore the full-stack in its entirety. We hope that AcousTools can become the standard software package for acoustic holography applications, reducing development efforts while expanding opportunities for innovation. We encourage the community to adopt AcousTools as a unified platform, which will lead to greater collaboration, interoperability, and accountability. 

\printbibliography

\end{document}